\documentclass[pre,aps,twocolumn,showpacs,showkeys]{revtex4}
\usepackage{graphicx}
\usepackage{bm}

\def\be{\begin{equation}}
\def\ee{\end{equation}}
\def\bea{\begin{eqnarray}}
\def\eea{\end{eqnarray}}

\def\pa{\partial}

\begin{document}

\title{Vibrational properties of amorphous silicon from tight-binding O(N) 
calculation}

\author{Parthapratim Biswas}
\affiliation{Debye Institute, Utrecht University, Princetonplein 5, 
3508 TA Utrecht, The Netherlands}
\email{P.Biswas@phys.uu.nl}


\begin{abstract}
We present an $O(N)$ algorithm to study the vibrational properties 
of amorphous silicon within the framework of tight-binding approach. 
The dynamical matrix elements have been evaluated numerically in the
harmonic approximation exploiting the short-range nature of the 
density matrix to calculate the vibrational density of states which 
is then compared with the same obtained from a standard O($N^4$) 
algorithm. For the purpose of illustration, an 1000-atom model is 
studied to calculate the localization properties of the vibrational 
eigenstates using the participation numbers calculation.
\end{abstract}

\maketitle 

In recent years there has been a tremendous progress in the 
development of efficient $O(N)$ methods for electronic structure 
calculation of materials. There exist a number of $O(N)$ methods 
which have been developed to calculate the total energy of the 
systems and the electronic 
forces~\cite{goedecker,ordejon,vanderbilt,stephen,baer,gillan}. 
Most of these algorithms utilize, in one way or another, the fact 
that the density matrix of the system to be studied is 
short-ranged~\cite{goedecker,baer}. As a result, calculation of 
local electronic properties are possible by introducing a localization 
volume surrounding the region of interest where the local electronic 
properties are to be studied. The size of the localization volume 
depends on the decaying nature of the density matrix and for systems 
with a clean band gap, e.g., semi-conductors it has been observed 
that the density matrix decays exponentially~\cite{goedecker,stephen,baer}. 
Hence it is possible to define a localization volume having a linear 
dimension of the order of localization length of the density 
matrix. Although this does not appear to be trivially true in 
metallic systems~\cite{note1}, the {\em locality} in electronic 
structure calculation does indeed exist irrespective of the 
nature of the system and was first utilized by Friedel~\cite{friedel} 
who introduced the term {\em local density of states}. Later 
Heine~\cite{heine} extended this idea of local electronic structure 
calculation of solids in real space in the context of Recursion 
method~\cite{heine,hhk} which has been further elaborated by 
Kohn~\cite{kohn} as the {\em principle of nearsightedness of 
equilibrium systems}. 

In this paper we present an efficient algorithm for the calculation 
of vibrational properties of amorphous silicon from a 
tight-binding~\cite{colombo, bowler} Hamiltonian exploiting the 
short-range nature of the density matrix. The motivation behind 
this work is the calculation of localization properties of the 
vibrational eigenstates of amorphous silicon. Recently, there has
 been a renewed interest in the study of the vibrational modes in 
disordered media~\cite{allen,elliot,allen2}. These modes can be 
classified as either extended or localized depending on the energy 
of the state. In $d=3$, the modes are expected to separate by sharp 
energy boundary, known as `mobility edge' which marks the `localized 
to delocalized' transition. The corresponding analog in the electronic 
system is well-known after Anderson's suggestion~\cite{anderson} 
that, unlike crystalline systems where electronic states are extended 
over the system, sufficient disorder can localize electronic states 
within the allowed energy bands. For vibrational eigenstates, however, 
the low frequency, long wavelength modes (sound) always exist. 
Garber et al.~\cite{allen2} studied an amorphous model of 4096 
silicon atoms and demonstrated that at the mobility edge the 
vibrational eigenstates decay exponentially and localization length 
diverges as a power law above the edge. Although we are not going 
to address these issues here, the  importance of the construction 
of dynamical matrix from tight-binding calculations and 
to study the nature of the vibrational states for a large, 
well-relaxed sample of amorphous silicon has its own merits. 

In the following, we illustrate our work in two steps: first, we 
calculate the approximate electronic forces using an $O(N)$ 
algorithm starting from a tight-binding Hamiltonian. This is then 
followed by a calculation of harmonic normal modes of vibrations by 
direct diagonalization of the dynamical matrix. The diagonalization 
of a matrix is an expensive $O(N^3)$ operation, but since the time 
required to construct the dynamical matrix elements using this 
$O(N)$ approach is much larger compared to a single diagonalization 
for the system sizes to be considered here, we make no attempt to 
obtain the density of states using any $O(N)$ algorithm. In addition, 
we also study the nature of the vibrational eigenstates for which 
it is necessary to compute the eigenvectors of the dynamical matrix 
for participation numbers calculation. For the density of states 
calculation only, however, one can use any one of the $O(N)$ 
algorithms available in the literature which include Maximum Entropy 
approach by Drabold and Sankey~\cite{drabold}, Recursion method of 
Haydock, Heine and Kelly~\cite{hhk} and Kernel Polynomial approximation 
by Silver and Roder~\cite{silver1,silver2} etc. For our present 
purpose, a single diagonalization is easily affordable for the calculation 
of spectral quantities whereas the construction of a dynamical 
matrix using the exact electronic forces by direct diagonalization 
is extremely inefficient and time consuming which scales as $O(N^4)$ 
making it a formidable task to go beyond 500 atoms. 
This direct diagonalization approach is not practical and feasible 
for very large system size ($\ge$ 10000 atoms). However, one 
can still obtain spectral density distribution using any one of the 
methods mentioned earlier and a few eigen vectors in appropriately 
chosen intervals using Lanzcos algorithm in an $O(N)$ way to study 
the nature of the states. 

Throughout this work, we have used the structural model generated by 
Barkema and Mousseau~\cite{barkema} as starting configurations which 
have been relaxed using the tight-binding Hamiltonian parameterized 
by Kwon et al.~\cite{kwon}.

The general form of the tight-binding Hamiltonian $H$ can be 
written as :
$
H = \sum_{i,L} \epsilon_{i,L} \vert {i,L} \rangle \langle {i,L}\vert
+ \sum_{i,L}{\sum^{'}_{i',L'}} V_{i,L;i',L'} \vert {i,L} \rangle \langle
{i',L'}\vert
\label{hamil-1}
$ 
and the corresponding total energy is $
E = 2\sum_{k}^{occup} \langle \Psi_k \vert H \vert \Psi_k \rangle + U_{r}
+ E_{at}\,N. 
$ 
$U_{r}$ stands for repulsive potential for the ion-ion interaction
which can be represented either by a sum of pair potentials or as in the 
embedded atom model approach~\cite{embed} as a functional of atomic 
density. $N$ is the total number of atoms and $E_{at}$ is a constant 
energy/atom. The electronic eigen function $|\Psi_k\rangle $ of 
$H$ can be expanded as a linear combination : $|\Psi_k\rangle = 
\sum_{i,l} b_{i,l}^k | \phi_{l,i}\rangle $ of basis functions 
$\{|\phi_{l,i}\rangle \}$. For amorphous silicon, the minimal basis 
functions consist of one $s$-electron and three $p$-electrons per atom. 

The dynamical matrix element between two atoms at sites $i$ and 
$j$ can be written as~\cite{drabold0}
$$
D_{i,\alpha}^{j,\beta} = \frac{1}{\sqrt{M_i M_j}} \left[\frac{F_n^{j,\beta} - 
F_o^{j,\beta}}{\Delta x_{i,\alpha}}\right] = \frac{1}{\sqrt{M_i M_j}} 
\left[\frac{\Delta F^{j,\beta}}{\Delta x_{i,\alpha}} \right],
$$
where $M_i, M_j$ are the atomic masses at the sites $i$ and $j$ 
respectively. $F_o^{j,\beta}, F_n^{j,\beta}$ are the forces acting 
on the atom at the site $j$ in the $\beta \,(= x, y,z)$ direction 
due to displacement $\Delta x_{i,\alpha}$ of the atom at the site 
$i$ in the $\alpha$ direction before and after the displacement 
respectively. A straight forward implementation~\cite{colombo,bowler} 
of the above procedure for obtaining dynamical matrix elements between 
all the atoms require at least $(3N+1)$ direct diagonalizations for a 
system of $N$ number of atoms. The total force acting on an atom 
(say $i$) can be written as,  

\bea
F_i & = & -\frac{\pa E}{\pa R_i} \nonumber \\
& = & -2\frac{\pa}{\pa R_i}\sum_k^{occup} <\Psi_k|H|\Psi_k>-\frac{\pa U_r}{\pa R_i} \nonumber
\eea

Using Hellman-Feynman theorem~\cite{feynman}, the electronic part 
can be written as : 
\bea
F^e_i & = & -2\frac{ \pa}{ \pa R_i} \sum_k \sum_{j,L} \sum_{j',L'} 
b^k_{j,L} b^k_{j' ,L'} < \phi_{j,L}|H|\phi_{j',L'} > \nonumber \\ 
      & = & -2\sum_k \sum_{j,L} \sum_{j',L'} b^k_{j,L} b^k_{j',L'}
<\phi_{j,L}|\frac{\pa H}{\pa R_i}|\phi_{j',L'}> \nonumber.  
\eea 

\begin{figure}
\includegraphics[width=3.4in,height=3.4in,angle=0]{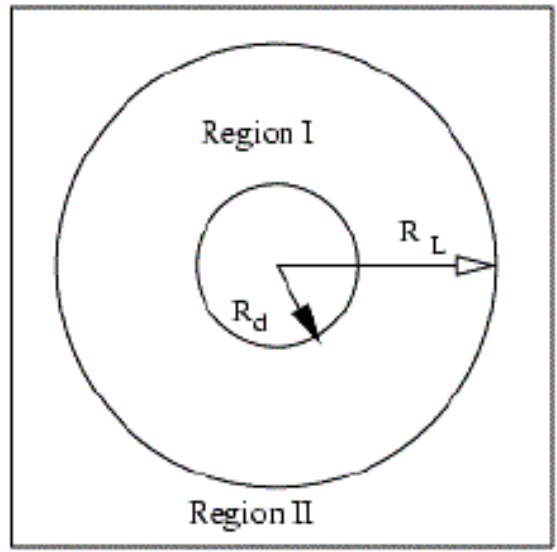}
\caption{\label{local} Partitioning of the entire space for the 
calculation of dynamical matrix elements between the atom at the 
center of the sphere and those inside the small sphere of radius 
$R_d$. The large sphere with radius $R_L$ corresponds to the 
localization volume (denoted by region I) for the atom at the center 
contributing to the forces on the atoms inside the small sphere.
}
\end{figure}

The calculation of eigen vectors (b's) is 
an $O(N^3)$ operation and for the entire system (containing 
$N$ atoms) the scaling goes as $O(N^4)$. For each dynamical matrix 
element calculation one has to calculate the electronic force 
before and after the displacement of the atom and this has to 
be repeated for each direction $\beta (=x,y,z)$. For $N \ge 
1000$, it is computationally overkill to construct the full 
dynamical matrix for the system. The calculation, however, can 
be performed in an $O(N)$ way by introducing a localization 
volume surrounding the atom on which the force is to be calculated 
as described below.

To calculate the electronic force using an $O(N)$ approach, 
we partition the entire system into regions I and II with the 
the atom as center as shown in figure \ref{local}. The large
sphere with radius $R_L$ corresponds to the localization volume 
for the central atom at $i$. The dynamical matrix elements between 
the atom $i$ at the center and those inside the small sphere 
of radius $R_d$ are to be constructed. Typically, $R_d$ can be 
chosen as second neighbor distance of the central atom or more. 
By varying the radius $R_d$ ($R_d < R_L)$, one can monitor the 
magnitude of the dynamical matrix elements and can check whether 
they affect the vibrational density of states or not. The far 
away the atoms are from the central atom, the less is the effect 
of displacement on the atom at $j$~\cite{note2}. 

\begin{figure}
\includegraphics[width=3.in,height=3.4in,angle=270]{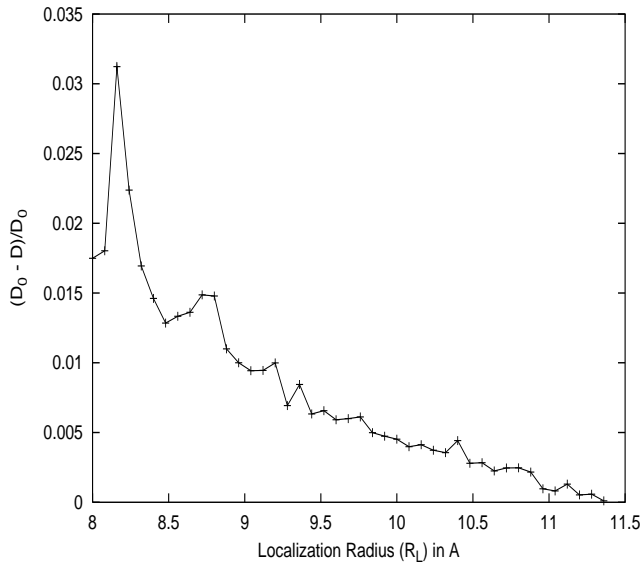}
\caption{\label{conv} 
The convergence behavior of dynamical matrix elements for different 
values of localization radius $R_L$. $D$ and $D_0$ are the approximate 
and exact dynamical matrix elements respectively for each of the $R_L$ 
values. For $R_L > $ 10\AA, the error is observed to be less than $1$ 
percent.}
\end{figure}

One can, in fact, introduce a cut-off for dynamical matrix 
elements while storing the elements. Here we have used a value 
of {3.9\AA} for $R_d$ and {10\AA} for $R_L$ which we find is 
sufficient for the purpose of our calculation. The difference 
of force on an atom $j$ due to the displacement of 
the atom $i$ at the center can be written as : 
\bea 
\Delta F^j &=& \left[F_{n,I}^j + F_{n,II}^{j}\right] - \left[ F_{o,I}^{j} +
F_{o,II}^{j}\right] \nonumber \\ 
&=& \left[F_{n,I}^{j} - F_{o,I}^{j}\right] + \left[ F_{n,II}^{j} - F_{o,II}^{j} 
\right] \nonumber \\
\Delta \tilde F^j & \approx & \left[F_{n,I}^{j} - F_{o,I}^{j}\right] \nonumber
\eea 
where $F^j_{o,I} (F^j_{o,II})$ and $F^j_{n,I} (F^j_{n,II})$ are 
the contribution to the total forces from the region I (region II) 
on the atom $j$ before and after the displacement of the atom $i$ 
at the center respectively. 
$\Delta F^j$ and $\Delta \tilde F^j$ are the exact and approximate 
difference of total forces at the atom $j$. Our $O(N)$ algorithm is 
based on the assertion that the approximate difference of total force 
can be obtained by neglecting the contribution from region II if 
the radius of the region I is greater than the localization length 
of the density matrix. This assertion can be verified simply by 
looking at the convergence behavior of the dynamical matrix elements 
with respect to the radius of the localization volume. The approximate 
difference of force can now be written as a sum of electronic and 
repulsive contribution : 
\bea 
\Delta \tilde F^j & = & \Delta \tilde F^{j,e} + \Delta \tilde F^{j,r} \nonumber \\
\Delta \tilde F^{j,e} & = & (F^{j,e}_{n,I} - F^{j,e}_{o,I}) \nonumber 
\eea 
\begin{figure}
\includegraphics[width=3.in,height=3.in,angle=270]{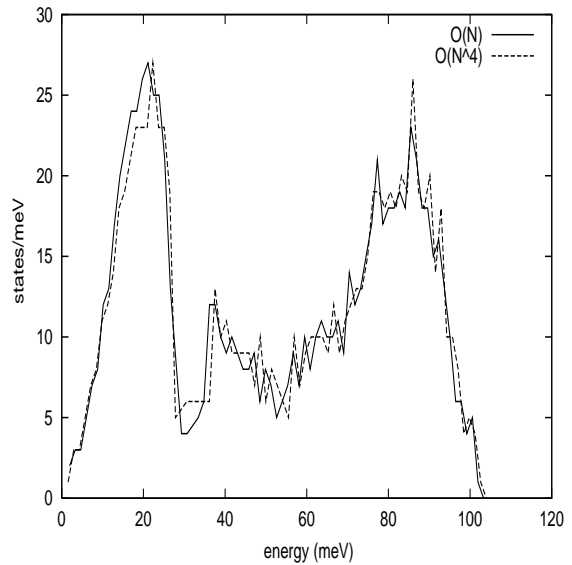}
\caption{\label{v500} The density of vibrational states for a system 
of 500-atom model calculated using the O(N) force algorithm described 
in the text. The same from a standard O($N^4$) force calculation is 
also plotted (denoted by a dotted line) for comparison. }
\end{figure}
with 
$$
F^{i,e}_{x,I}=-2\sum_k\!\sum_{jL,j'L'}\!\!(b^{k,I}_{j,L})^x (b^{k,I}_{j',L'})^x <\phi_{j,L}|\frac{\pa H^x_{I}}{\pa R_i}|\phi_{j',L'}> \nonumber 
$$
where $H^x_I$ is the Hamiltonian of the region I before ($x=o$) and 
after the displacement($x=n$) of the atom at the center. 

We have observed that a localization volume of radius {10\AA}, 
typically containing 200 to 225 atoms, gives a very good estimate 
of $\Delta \tilde F^j$ for calculation of dynamical matrix 
elements. If we neglect contributions from region II, the force 
calculation on a single atom is an $O(1)$ operation as the size 
of region I is independent of the total system size and the full 
dynamical matrix construction is an $O(N)$ operation for $N$ 
atoms. Since the dimension of the truncated Hamiltonian matrices 
$H^I$ are much smaller than the Hamiltonian matrix $H$ of the total 
system, the electronic forces can be calculated by direct diagonalization 
of the truncated Hamiltonian matrix $H_I$ which is also independent 
of the system size. In practice, the accuracy of $\Delta \tilde F^j$ 
depends on two factors -- the step size of displacement and the 
volume of region I which is again related with the size of band-gap. 
The accuracy can be improved in a controlled manner by (a) choosing 
a small displacement which is a necessary condition for the harmonic 
approximation to be valid and, (b) increasing the radius of the 
localization volume.
\noindent 
We would like to mention that a similar but not identical approach 
based on the use of localized Wannier-like functions and truncating 
the dynamical matrix beyond a cutoff was reported by Ordejon et 
al.~\cite{wan}. 

\begin{figure}
\includegraphics[width=3.in,height=3.in,angle=270]{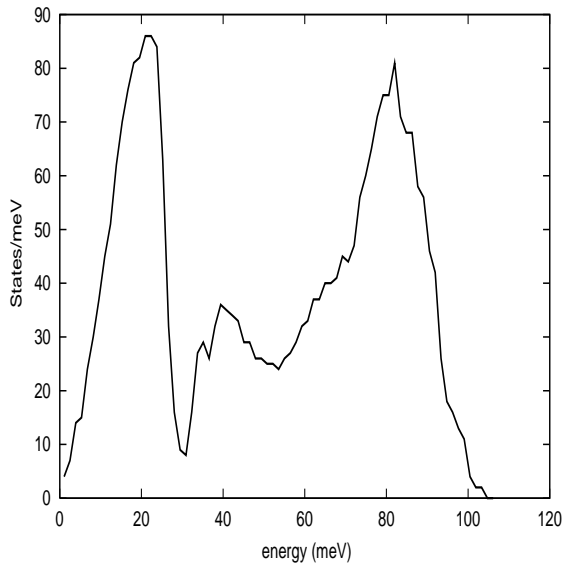}
\caption{\label{v1000} The vibrational density of states for 
a system of 1000-atom obtained using the $O(N)$ force algorithm 
described in the text.}
\end{figure}

In figure~\ref{conv} we have presented convergence behavior of 
dynamical matrix elements with respect to different localization 
radius $R_L$. The estimate of approximate electronic forces and 
hence the dynamical matrix elements are found to be very good 
and the percentage error is observed to be as low as 0.5 for 
$R_L \ge$ {10\AA}. The results of our calculations are presented 
in the figure~\ref{v500} where we have plotted the density of 
vibrational states obtained from a 500-atom model. For the purpose 
of comparison, we have also plotted the density of states using 
the full Hamiltonian matrix by a standard $O(N^4)$ algorithm. It 
is quite clear from the figure~\ref{v500} that the quality of 
(electronic) forces obtained from the $O(N)$ algorithm is very 
good giving the same vibrational density of states as one would 
expect using the full Hamiltonian matrix. We should mention that 
to generate the dynamical matrix elements for this 500-atom model 
using standard $O(N^4)$ force algorithm took 80 hours in contrast 
to 13.5 hours using the $O(N)$ algorithm in a DEC alpha (667 MHz) 
workstation. 

Having established that the dynamical matrix obtained in this 
$O(N)$ way is sufficiently accurate, we now study the nature 
of the vibrational states for a model of 1000-atom. The configuration 
is first relaxed using the tight-binding Monte Carlo calculation 
as outlined in the Ref.~\cite{biswas}. The vibrational density 
of states of the Monte Carlo relaxed structure is shown in the 
figure~\ref{v1000} which has been obtained by direct diagonalization 
of the dynamical matrix. Since the dynamical matrix construction 
for this 1000-atom model took approximately 27 hours, we therefore 
avoid to use any one of the $O(N)$ algorithm mentioned in the 
Refs.~\cite{drabold,silver1,silver2} for the calculation of 
vibrational density of states (VDOS). 
The VDOS in figure \ref{v1000} clearly shows two peaks :  a low 
energy acoustical peak at 20 meV and a high energy optical peak 
at 80 meV. 
\begin{figure}
\includegraphics[width=3in,height=3in,angle=270]{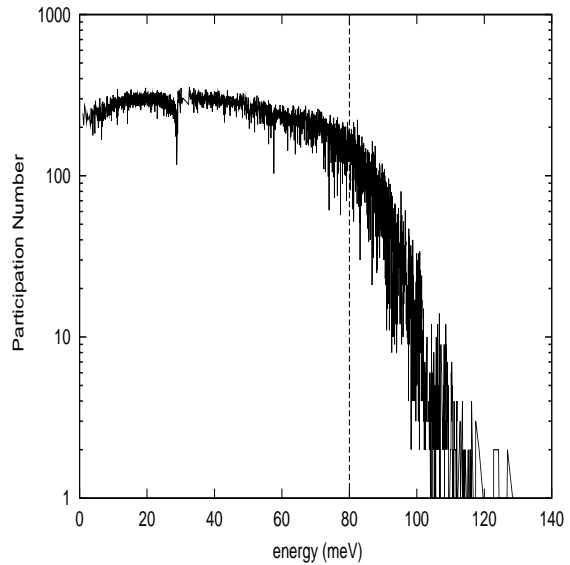}
\caption{\label{pr1000} The participation numbers (in log scale) 
for a model of 1000-atom with energy. The number drops near around 
80 meV showing the localized character of the eigenstates which is 
marked by a vertical line. }
\end{figure}
The ratio of the height of two peaks are of the same 
order as observed in the experimental data obtained by Kamitakahara 
et al.~\cite{kam}. The optical peak is, however, at higher energy 
compared to the experimental value and the value obtained from 
empirical potential calculations. Such a high energy optical 
peak has been also observed in the calculation of Klein et 
al.~\cite{klein} who studied hydrogenated amorphous silicon using 
tight-binding molecular dynamics. We believe that this discrepancy
 in the position of the high energy optical peak is associated 
with the repulsive part of the potential used in this calculation. 
The potential produces the correct electronic density of states 
within the accuracy of a fraction of an eV; however, for the 
vibrational density of states calculation an accuracy of the order 
of few meVs is required to produce the peaks at right positions. 

In order to get an idea about the nature of the vibrational states, 
we need to calculate the participation numbers from the eigenvectors 
of the dynamical matrix. Since the eigenvectors calculation is a 
memory as well as CPU intensive operation, it is not a very suitable 
tool for dealing with very large system sizes. We therefore restrict 
ourselves to an 1000-atom model for this purpose. The participation 
numbers give a measure of the degree of localization of the wave 
functions to be studied. Let $\phi_j$ denotes the amplitude of the 
$j$-th normalized eigenvector. A simple measure of the number of 
sites participated in contributing the eigenstate gives the participation 
number $P_N(j)$ and is defined as 
$P_N^{-1}=\sum_{i}|\phi_j(i)|^4$. 
A complete localized state therefore corresponds to $P_N = 1 $, whereas 
for an extended state $P_N = N$. In figure \ref{pr1000}, we have plotted 
the participation numbers with energy in meV.  In the low energy region 
the value of the participation number fluctuates around 300. The drop of 
the participation numbers around 80 meV is actually somewhat higher than 
that was observed in the work of Allen et al.~\cite{allen} who found a 
value of around 70 meV. This difference is, however, consistent with what 
we have observed in figure~\ref{v1000} where the optical peak appears at 
somewhat higher energy. In spite of this difference, we find that the 
general trend of the participation numbers with energy (meV) follows 
the same course as was observed in Ref.~\cite{allen}. For example, 
a small dip at 30 meV in the figure~\ref{pr1000} has been also observed in 
Ref.~\cite{allen}. A closer inspection of the localization properties using 
the participation numbers requires a much larger system size to be studied 
for better statistics, particularly, in the high energy region where a 
delocalized - localized transition is expected to occur from the low energy 
side. The tight-binding calculation for a model of 10~000-atom is in progress 
and in a future communication we will address the energy-level-spacing 
distribution and eigen function statistics of the spectrum. 

In conclusion, we have presented an $O(N)$ algorithm for the calculation 
electronic force for amorphous semiconductors from tight-binding Hamiltonian 
exploiting the short-range nature of the density matrix. 
This algorithm has been used to construct the dynamical matrix for 
a 500-atom model of amorphous silicon. The vibrational density of 
states calculated from this $O(N)$ dynamical matrix has been observed 
to be in excellent agreement with the same obtained from a standard 
$O(N^4)$ algorithm. 
Furthermore, we have used this algorithm to study the nature of the 
vibrational states for a model of 1000-atom amorphous silicon, the 
dynamical matrix of which is otherwise difficult to construct in 
the tight-binding approach. The participation numbers calculation 
suggests that the results of this tight-binding calculation are in 
agreement with the results based on empirical model potentials.

The author acknowledges G.~T.~Barkema of Institute for Theoretical 
Physics, Utrecht University for providing the coordinates of amorphous 
silicon samples which have been used as starting configurations in the 
present calculations. The author also thanks David Drabold of Ohio 
University for a careful reading of the revised manuscript. The work 
has been supported by NWO within the priority program ``Solar Cells in 
the 21$^{st}$ Century''.


\begin{thebibliography}{*99}
\bibitem{note2}
This is not entirely true in presence of a long-range interaction 
when the locality principle can not be applied trivially. In this 
respect, our method can not be applied to systems where the long-range 
contribution to electronic force is significant.
\end{thebibliography}
\end{document}